\documentclass[aps,prb]{revtex4}   
\usepackage{graphicx}  

\begin{document}

\title{Scouting the spectrum for interstellar travellers}
\author{Juan Carlos Garcia-Escartin}
\author{Pedro Chamorro-Posada}
\affiliation{Universidad de Valladolid, Dpto. Teor\'ia de la Se\~{n}al e Ing. Telem\'atica, Paseo Bel\'en n$^o$ 15, 47011 Valladolid, Spain.\\
Corresponding author's e-email: juagar@tel.uva.es}
\date{\today}

\begin{abstract}
Advanced civilizations capable of interstellar travel, if they exist, are likely to have advanced propulsion methods. Spaceships moving at high speeds would leave a particular signature which could be detected from Earth. We propose a search based on the properties of light reflecting from objects travelling at relativistic speeds. Based on the same principles, we also propose a simple interstellar beacon with a solar sail. 
\end{abstract}

\maketitle

\section{Indicators of advanced civilizations}
\label{indicators}
The search for life outside the Earth faces several challenges. The vastness  of space makes it difficult to single out promising candidate regions and, due to the large distances involved, indirect observation techniques must be used. The choice of adequate indicators is crucial. 

The detection of life usually focuses on finding planets and biological markers, such as water or the presence of carbon or oxygen (Sagan et al., 1993; Leg\'er et al., 1996). The search for intelligent life concentrates mostly on listening to possible transmissions, like in the SETI programme (Tarter, 2001), or looking for other pointers to civilization, such as artificial illumination (Loeb and Turner, 2011) or probes (Freitas, 1983). 

Interstellar ships of different types could leave traces of their presence. There are proposals to look for the residues of propulsion with nuclear or antimatter drives. Those ships would leave a characteristic signature in the $\gamma$ and X ray bands, among others (Zubrin, 1995). We suggest an additional indicator for the detection of hypothetical advanced civilizations capable of interstellar travel. We study the light reflecting from high speed starships, which, in principle, does not depend on the concrete propulsion method but on the final velocity.

While there is no guarantee that there is life in any other place of the Universe, let alone extraterrestrial civilizations, it is reasonable to suppose that advanced intelligent civilizations, or, at least, a fraction of them, would be interested in space travel. They could undertake space travel for exploration, the exploitation of resources of nearby stellar systems or survival when their home star undergoes changes that can render their home world uninhabitable (Matloff and Pazmino, 1997). The universal speed limit imposed by Relativity and the long distances between stellar systems makes finding advanced methods of propulsion a priority for space exploration. We propose looking for the signature of artificial objects moving close to the speed of light as a way to identify intelligent life. We can explore the spectrum looking for Doppler shifts caused by reflection on high speed objects.

In Section \ref{REP} we define the range of velocities of natural objects and the range of extraordinary propulsion we suggest exploring. Section \ref{RelMirr} describes reflection from relativistic mirrors. Sections \ref{light} and \ref{first} study the light reflected from interstellar ships under different circumstances. Section \ref{Sailbeacon} suggests a way to build an interstellar beacon with a solar sail, based on the properties of relativistic mirrors. Finally, Section \ref{prospects} discusses the possibilities and limitations of this search approach.

\section{Extraordinary propulsion}
\label{REP}
\begin{figure}[ht!]
\centering
\includegraphics{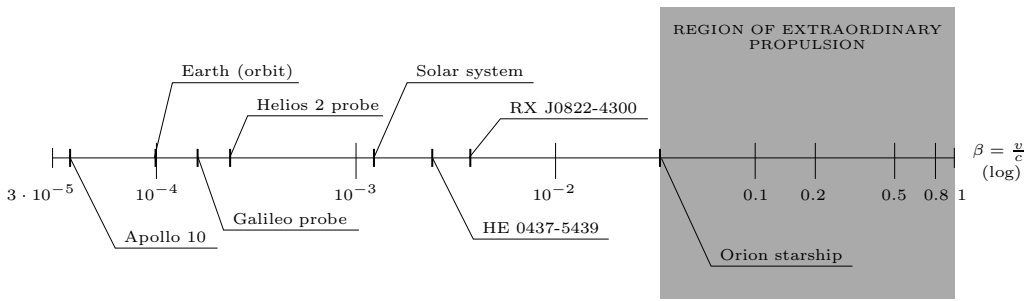}
\caption{Scales of speed with respect to the speed of light in vacuum (logarithmic scale). The fastest man-made objects are in the range of velocities from $10^{-5}c$ to $10^{-3}c$. Examples are the fastest manned ship, Apollo 10 on entry  (NASA, 1969), the Galileo probe during its descent into Jupiter (NASA, 2003) and the solar probe Helios 2, with its velocity estimated at its periapsis at 0.29 Astronomical Units from the Sun (Freeman, 1998). For comparison, we have included the average speed of Earth during its orbit around the Sun (Cox, 2000) and the motion of the Solar System with respect to the cosmic microwave background frame (Hinshaw et al., 2009). The fastest natural objects, like hypervelocity star HE 0437-5439  (Brown et al., 2010) and neutron star RX J0822-4300 (Hui and Becker, 2006), move in the scale of $10^{-3}c$-$10^{-2}c$. We define a region of extraordinary propulsion (REP) for speeds which would point to an artificial object. The REP starts at the estimated speed for the nuclear propulsion Orion ship (Dyson, 1968), which could be built with present human technology.\label{speeds}} 
\end{figure}

Few natural objects achieve the relativistic speeds we can expect from interstellar travellers. Only massless or very light particles come close to the speed of light. Objects of a certain size do not achieve on their own more than a small fraction of the speed of light in vacuum, $c$. Figure $\ref{speeds}$ shows the maximum velocities which have been observed from different stellar objects and compares them to the fastest human-made ships.

Of particular interest is the velocity range we call region of extraordinary propulsion, REP. This region encompasses the range of speeds which any intelligent civilization should aspire to if they want to explore, at least, their nearest stellar systems. We choose as the beginning of this range the estimated speed of the nuclear propulsion ship proposed in the Orion project, with a speed  of $3.3\% c$ (Dyson, 1968). This lowest velocity is achievable with human technology using nuclear blasts from fission bombs. Nevertheless, it would still be a formidable challenge in engineering and economical terms. The upper range can include any advanced propulsion method either proposed or yet unknown. Examples could be advanced nuclear propulsion, such as that of the Daedalus, Icarus or Longshot projects (Long et al., 2011), antimatter drives (Forward, 1982), Bussard ramjets (Bussard, 1960), solar sails (Vulpetti et al.,  2008) or more exotic means (see Matloff (2000) for a good survey on propulsion methods). Most of them can achieve speeds at least in the range of $0.1c$, depending on the particular implementation. The only requisite for all these propulsion systems is they respect the relativistic speed limit and travel through normal space (no wormholes or other short cuts through a supposed hyperspace), which is, nevertheless, a likely restriction. Even if wormholes are available, parts of the total distance should be covered with the usual restrictions. Most proposals short of teleportation require an additional normal space stage, possibly at high speed. 

Motion at those speeds would point to artificial objects. Even the slowest from these mechanisms could provide a velocity above the observed velocities for natural objects. Natural velocities have their record at the speed of the radio-quiet neutron star RX J0822-4300 in the supernova remnant Puppis-A, which is estimated to be between $0.25\%c$ and $0.5\%c$ (Hui and Becker, 2006). Even if we compose the speed of two of these stars moving away from each other, we would still be one about one order of magnitude below the Region of Extraordinary Propulsion. Hypothetical objects, like hypervelocity planets, could enter the REP. If one star from a binary system is captured by a massive black hole while its companion is ejected as a hypervelocity star, a planet from the binary system could, in some rare situations, reach a speed as high as $0.1c$ (Gingsburg et al., 2012). In any case, even if the existence of such objects is confirmed, their number would be small and their locations limited. We can assume, without much error, that no natural object can be found in the REP. If an object is found at that speeds, we can later check for possible natural origins.  

Our method of detection is based on the observation of the Doppler shifts in light reflected from moving ships. The shifts in the spectrum of the source of illumination, under certain conditions, can be greater than could be expected from any natural object and would provide a detectable signature of space travel. For high speeds, there are situations in which relativistic effects amplify the reflected light, allowing for the detection of relatively small objects.

\section{Relativistic mirrors}
\label{RelMirr}
Our detection proposal rests on the properties of relativistic mirrors. For simplicity, we describe first a mirror moving in the direction of the light it reflects. The mirror has a speed $\beta=v/c$, normalized to the speed of light, which can be positive (the mirror moves towards the source) or negative (the mirror moves away from the source). 

In reflection from a relativistic mirror, we see a double Doppler shift. If the source emits photons of frequency $f$ the mirror sees them at frequency $f'=\alpha f$ where is $\alpha$ the relativistic Doppler factor $\sqrt{\frac{1+\beta}{1-\beta}}$. The reflected signal, at frequency $f'$, but going in the opposite direction, will appear in the frame of the source at a frequency $f''=\alpha f'$. The global effect, for a transmitter and a receiver in the same frame, is a Doppler shift by a factor $\alpha^2$. This result was already discussed by Einstein in his first Special Relativity paper (Einstein, 1905).

We are assuming a perfect mirror with a momentum much higher than the photons. There is a small correction if we take into account the mirror's recoil on reflection. A single photon reflecting from a mirror of mass $M$ with a Lorentz factor $\gamma=\frac{1}{\sqrt{1-\beta^2}}$ satisfies relativistic conservation of energy and momentum equations:
\begin{eqnarray}
\gamma M+hf&=&\gamma' M +h f',\label{consE}\\
\gamma \beta M - hf&=&\gamma' \beta' M +h f'. \label{consP}
\end{eqnarray}
The signs correspond to a photon reaching the mirror in the direction opposite from the positive velocities $\beta$. If we add and subtract Equations (\ref{consP}) and (\ref{consE}) we see that:
\begin{equation}
f''=\alpha \alpha' f
\end{equation}
where, $\alpha$ and $\alpha'$ are the Doppler factors corresponding to the velocities $\beta$ and $\beta'$ of the mirror's frame before and after the reflection. The speed is reduced as an effect of the radiation pressure. The mirror recoils after reflecting the photon. The final frequency can also be expressed as
\begin{equation}
f''=\alpha^2\frac{M}{M+2hf\alpha}f. 
\end{equation}
We can check that, for a large mirror ($M\gg 2h f \alpha$), we can essentially ignore recoil as far as the Doppler effect on the reflected light is concerned. If we have many photons, the correction can be ignored if the mass of the mirror is much greater than twice the energy of all the photons in the mirror's frame.

For simplicity, we also assume the receiver (us, on Earth or a space station) is in the same frame as the light source, usually a star. If we consider the Doppler factor between the frames of the ship (small s) and the star (capital S), $\alpha_{Ss}$, and the Doppler factor between the frames of the ship and Earth (E), $\alpha_{sE}$, we can also write a more general total Doppler shift $f''=\alpha_{Ss}\alpha_{sE}f$. These factors can include different lines of motion. The Doppler factor considering any direction of motion  with an angle $\theta$ between the directions of motion and light propagation is
\begin{equation}
\alpha=\frac{1-\beta\cos{\theta}}{\sqrt{1-\beta^2}}.
\end{equation}
Even for an angle $\theta=\frac{\pi}{2}$ there is transverse Doppler factor $\frac{1}{\sqrt{1-\beta^2}}$.

Apart from the frequency shifts, there are important effects on the signal's amplitude and spatial extension (McKinley, 1979). The signal's power is multiplied by $\alpha^4$. If we have a photon flux of $\Phi$ photons per second, each with energy $hf$, we reflect those photons with energy $hf''=h\alpha^2 f$. The photon flux changes due to time dilation effects and becomes $\alpha^2\Phi$. An interval of a certain length in the source frame appears shorter after reflection. A $T$ seconds period lasts $T/\alpha$ seconds in the mirror's frame and, as seen back in the source's frame, is $T/\alpha^2$ seconds long. This symmetry is slightly counterintuitive, but we can see it is necessary or, otherwise, there would be a preferred frame. 

There is one final effect regarding the direction of the signals. The angles of arrival are also transformed. A photon coming from an angle $\theta$ is reflected at an angle $\theta''$ so that 
\begin{equation}
\cos\theta'' = \frac{(1+\beta^2)\cos\theta-2\beta}{1-2\beta\cos\theta+\beta^2}.
\end{equation}
The net effect is sometimes called Doppler beaming or headlight effect. Signals which are redshifted ($\alpha<1$) are also broadened during the reflection. Signals which are blueshifted ($\alpha>1$) and amplified are also concentrated in the direction of the motion. A ship which collects all the photons in a solid angle $\Omega$, when in motion, picks up the photons contained in a solid angle $\alpha^2\Omega$ of the emitter's frame (McKinley, 1980). The total number of photons becomes $\alpha^2$ times greater.

This Doppler beaming and the increase of energy in the moving frame has already been discussed for relativistic starships. A ship approaching the speed of light would eventually see all the light in front of it as coming from a single point (Lagoute and Devoust, 1994; McKinley and Doherty, 1978). Furthermore, the radiation would turn to the $\gamma$-ray band and would appear more energetic as the ship comes closer to $c$.

Any interstellar ship at high speed must deal with this radiation bath in order to survive the journey (Semyonov, 2008). The travellers could try to harness some of this energy to power the ship's systems, but a good fraction of it should probably be deflected. Light reflected from such a ship would suffer twice the effects of beaming, blue shift and time compression. The final receiver sees a total scaling factor of $\alpha^8$. This gives us an opportunity for detection.

\section{Light from interstellar ships}
\label{light}
Most interstellar ships will reflect at least part of the electromagnetic radiation they receive from their environment, if only to avoid being burnt. Some ships can be better reflectors than others. Solar sails are particularly suitable for this method of detection as they are, in essence, large mirrors moving at relativistic speeds. The spectral nature of the reflected light will depend on the stellar environment of the ship. We propose looking for displaced versions of the different stellar spectral types. Reflection will hardly be uniform in the whole spectral range, but an initial search could look for good, wideband, almost perfect  mirrors and then refine from there. 

This method contrasts with previous detection techniques in that it is not proposed for a particular spectral band. The frequency band of the reflected signals depends on the speed of the ship and the angle of observation. This is both a problem and an advantage. The available exploration range is  possibly too large to be useful, but it has the advantage that measurements taken for other purposes can be searched for frequency displaced spectral patterns. Even if radiation at any concrete frequency can be small, finding correlations between different parts of the spectrum can help to single out candidate regions of space for a detailed observation.

\section{First explorations}
\label{first}

\begin{figure}[ht!]
\centering
\includegraphics[scale=0.5]{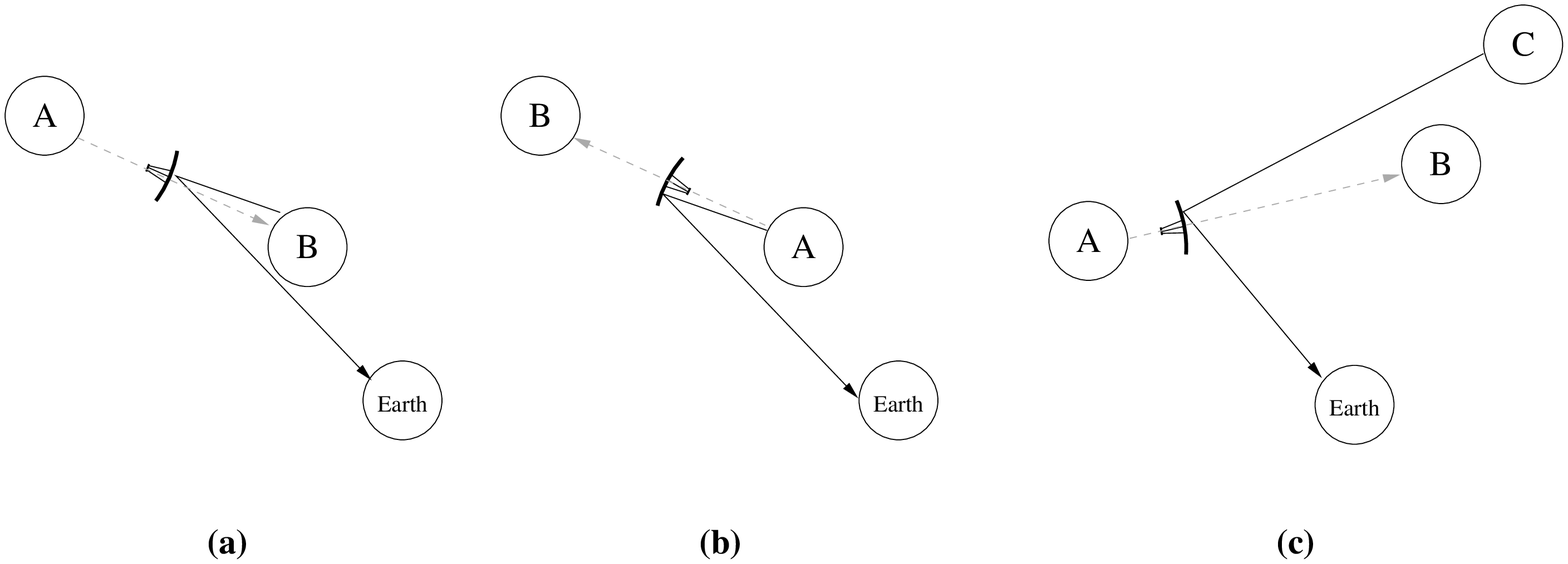}
\caption{Earth's position with respect to the ship's trajectory. {\bf(a)} Earth receives the light from the destination star reflected from an approaching ship. {\bf(b)} Earth receives the light from the origin star reflected from an outbound ship. {\bf(c)} Earth receives the light from a third star, which is reflected from the ship at an angle.\label{lines}} 
\end{figure}

In any extraordinary propulsion detection program we need to narrow down the possible candidate regions of space. 
There are certain routes for which we can expect more ships. In particular, due to the large distances between stellar systems and the speed of light limit, many interstellar journeys should occur between close locations. Even in this case, the journeys can span several years. It is reasonable to suppose that the travellers will take the straightest possible line from point A to point B (usually stars or objects in their immediate vicinity). The line will not be perfect. Many propulsion techniques, like gravity pull assisted ships or solar sails, restrict the available trajectories. Nevertheless, we can consider a direct or an almost direct trajectory in a first approximation. 
In a preliminary search, we should look for ships in a journey between two stars almost aligned with Earth. For an origin point A and a destination B, there are different possible positions for the A-B line with respect to Earth (see Figure \ref{lines}). 

We propose focusing on the cases where the observation point on Earth is almost in line with A and B, with only a small angle to avoid the closest star eclipsing the ship. There are two reasons for this choice. 

The first reason is related to the strength of the light reaching the ship. Most of the light reflecting from the ship, or, at least, the light with the highest power, will come either from the origin or the destination star. We are assuming B has been chosen because it is the nearest star to A. Depending on the stage of the journey, the star closest to the ship will be either A or B. Unless there is a third star, C, much brighter than both A and B and still reasonably near to the ship's trajectory, the strongest light reaching the ship comes from A or from B. Exceptions to this heuristic rule can be analysed in a more detailed search.

Secondly, the most extreme Doppler factors appear for objects moving in the same direction as the light. For high speed objects, there are two primary regions of interest. In the first region ($\alpha\ll 1$) the reflected light would be quite faint but covers a large region of space due to relativistic beaming. In the second region ($\alpha\gg 1$), we have a focused, energetic signal. Light coming from the direction of motion seems more promising for a preliminary exploration. 
We can discard reflection with a high angle in initial searches. The Doppler amplification factor will be smaller and the original signal is likely to come from a more distant star. In the first search attempts, we can choose stars that, besides being almost in line with Earth, are no more than a few light years apart. Larger separations pose technological challenges that might be beyond the capabilities of some of the civilizations for which a shorter journey is feasible. 
This approach has some difficulties. The amount of reflected light can be exceedingly small. A ship moving from A to B would be easier to spot, but photons reflected, say from a photon sail pushed by B in the direction of A, could also be detected. The only scenario in which we can expect a good detection would be for a ship coming directly towards Earth. A float of space invaders approaching Earth could be easily detected looking for replicas of the sun's spectrum at higher frequencies. For less fantastic detection situations, the reflected signals will not have maximum power on Earth's direction. Still, there are many star pairs which could be probed.	

The effects from motion combine to our benefit in the case of a ship approaching Earth. The $\alpha^8$ factor in the received power is then maximized. We also have a maximum spectral shift $\alpha^2$. When compared to the same ship at rest with respect to the star and Earth's common frame, we have a smaller apparent bolometric amplitude. Considering all the light in all the spectrum, we have a change in magnitude around $\Delta m =-20\log_{10}\alpha$. An approaching ship looks brighter. Take for instance the solar sail described by Matloff and Pazmino (1997). A perfectly reflective disc-shaped solar sail of radius 1000 km at one astronomical unit from Pollux would present an apparent magnitude 26 at Earth. If the same ship is moving at a speed $0.25c$, the apparent magnitude becomes 23.78. In this example, the spectrum is shifted by a factor of 1.66 and we are roughly in the same frequency band. In general, we use bolometric magnitudes to avoid the reference frequency filters which appear in the most usual magnitude comparisons. 

Changes in magnitude can go from a modest $\Delta m=-0.287$ at the low end of the region of extraordinary propulsion (with a nuclear ship moving at 3.3\% the speed of light) to important brightness increases, like $\Delta m=-6$ and $\Delta m=-12.78$, for the higher end (speeds $\beta=0.6$ and $\beta=0.9$). For all the speeds in the region of extraordinary propulsion, the frequency shift can be readily appreciated in a spectrometer. The absorption lines in the stellar spectra would be noticeably displaced. 

There is a limiting factor to the visibility of the ship. Accelerating a massive object is harder. Faster ships are likely to be smaller and will have a smaller reflective area.  It is the faintest objects which experiment the greatest magnitude increase. 

There are some technical details that can be fine-tuned later. Reflection from the ship will not be perfect at all the frequencies (but has to be high at high speeds unless the high energy incoming light can be somehow else deflected or dealt with). When pointing our telescopes we should likewise consider the angle of reflection. Snell law does not describe well relativistic mirrors. There is a   correction, more important as the speed increases. 

\begin{figure}[ht!]
\centering
\includegraphics[scale=0.5]{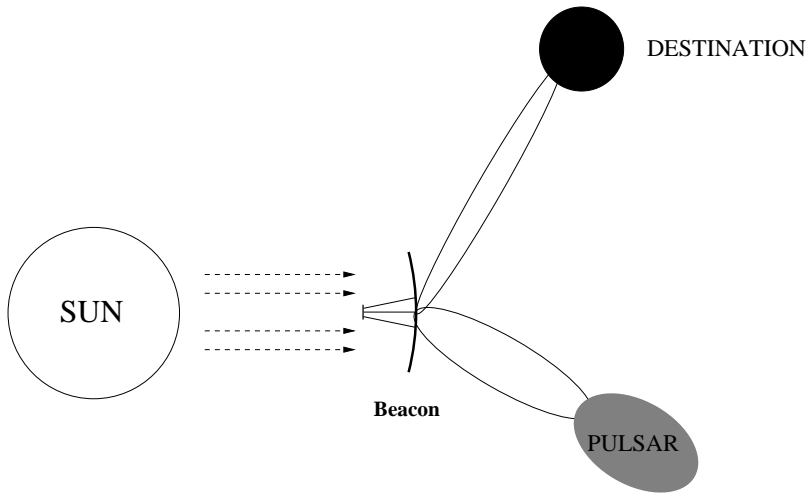}
\caption{Solar sail beacon. We can intentionally point fast reflective objects to candidate star systems. The sail plays the role of the mirror in a cosmic heliograph. If we want the signal to stand out among all the data from the Universe, we should reflect some interesting signal like a pulsar. The deviations from the expected frequency band are a clear indication of a relativistic mirror and, therefore, the presence of intelligent beings.  \label{beacon}} 
\end{figure}

\section{The solar sail beacon}
\label{Sailbeacon}

Civilizations with no access to interstellar travel can also take advantage of the extreme Doppler shifts in the region of extraordinary propulsion to broadcast their presence to other intelligent species. 

We propose a simple beacon based on a solar sail (Figure \ref{beacon}). We can launch a solar sail that reflects some interesting cosmic event, like a pulsar.  Pulsars can catch the attention of an advanced civilization with an interest in astronomy and have been proposed before as a cosmic common ground when searching for intelligent life (Edmondson and Stevens, 2003). An alien astronomer would notice two curious features in the reflected light. The first would be the evident Doppler shift, well beyond what can be expected from natural motion. The second would be its origin, close to the Sun. If we were to find such a displaced signal close to a star, we can infer the intentional presence of a cosmic lighthouse. 

Moreover, we can add modulation. If we include a rudder in the sail, we can direct the reflected signal in and out any desired stellar system. 
We can send alternate on and off pulses to likely candidate planetary systems. 

In contrast to a complete ship, the beacon does not need to carry a useful payload and must not necessarily withstand the harsh conditions of deep space travel for a long time. If the beacon is destroyed in the middle of the journey, it still serves its purpose. 

The construction of this kind of beacon can be a near term project for Humanity. Apart from sending a signal to the stars, it would be an ideal test bed  for solar propulsion technologies.  

\section{Prospects}
\label{prospects}

We have presented a new way to detect intelligent life. Massive objects moving at relativistic speeds do not appear naturally in the Universe as we know it. If we find the extreme Doppler shifts associated with reflection from such fast objects, it would be a strong evidence for an artificial origin. 

Interstellar ships could be detected with this method. For a ship moving towards Earth, all the relativistic effects combine to increase its visibility. The routes between close stars in line with Earth offer the best chance of detection. 

The same principles allow us to build interstellar beacons. A solar sail can work as a signalling mirror which directs light to any stellar system of choice. 

The proposed search programme has some limitations. As in all intelligent life search projects, there remains the distinct possibility that there are no civilizations sharing our time and space neighbourhood. Nevertheless, a detailed search could reveal yet unknown fast cosmic objects, like, maybe, rogue planets ejected when their stars are captured by a black hole. 

Interstellar travel could also be beyond the reach of most lifeforms. They could, instead, travel by proxy sending small robotic ships, like Bracewell probes (Bracewell, 1960). A smaller version of a solar sail pushed by a maser, the Starwisp, has also been suggested as an interstellar probe (Forward, 1985). These ships would be extremely small, but, precisely for that reason, they could achieve higher speeds. Other kinds of small ships can also appear. An example are repeaters. The same relativistic effects that make detection easier or more difficult depending on the direction of travel, affect the communication between a ship and its planet of origin. An outgoing ship sends and receives data at a smaller rate. A chain of small decode and resend repeaters moving at intermediate speeds would provide a ladder of frames for which good communication is restored (Garcia-Escartin and Chamorro-Posada, 2011). If these repeaters exist, we would have ships generating reflected replicas of the spectrum of the destination star at different frequency bands. These are just some examples of possible small ships. 

The first steps of a long term observation project can start now. We can compile a list of pairs of stars which are reasonably close to each other (for instance, no further than ten light years apart) and, at the same time, are in good alignment with Earth, allowing for a small deviation to avoid the origin and destination stars concealing any existing reflected signal. This list has already been started in previous interstellar migration detection proposals (Matloff and Pazmino, 1997). The list can be compiled taking  as a reference the habitable star catalogue, the HabCat (Turnbull and Tarter, 2003a, 2003b), so that we can focus on stars with the most potential for life.

The range of speeds we consider likely determines the frequency bands we should observe. In a first approximation, we can look into bands that should be empty. The less crowded frequency bands close to the usual stellar spectra are clear first candidates. Detection can be focused on the search for compressed or expanded replicas of the spectra of the origin and destination stars. These observations can have a relatively low cost. Many existing telescopes are sensitive enough to detect large ships in the higher velocity range. Better propulsion, corresponding to more technologically advanced civilizations, would be easier to detect. Finding ships at more plausible speeds will be more problematic, but not impossible. 
	If we search for relativistic beacons, we just need to look in the proximity of selected stars for strange sources of light. This is simpler than identifying interstellar routes between stars, but assumes an intentional emission in the direction of Earth. By focusing on the detection of light reflected from ships, we can keep the assumptions to a minimum. We do not need to suppose any intention to communicate or second-guess alien psychology to deduce the possible frequency bands and encodings of a message. The signal will also be independent from technology. The concrete propulsion system used in the ship is not important, only the final velocity. We only suppose a star-faring race with ships moving at relativistic speeds. The ships must be reflective, which, at those speeds, is a likely physical restriction. 

We can start combing existing data for a simple search. While the whole observation programme has no guarantee of success, a preliminary exploration has a low cost and it is worth at least an attempt. Even if no civilization is detected, we will be able to find or preliminary discard the presence of reflective fast objects near the stars, increasing our knowledge of the Universe around us.

\section*{Acknowledgements}
This work has been funded by project VA-342B11-2  (Junta de Castilla y León) and TEC2010-21303-C04-04.

\section*{References}

Bracewell, R. N. (1960) Communications from superior galactic
communities, \textit{Nature,} 186, 670--671.\vspace{0.5ex}

Brown, W.R., Anderson, J., Gnedin, O.Y., Bond, H.E., Geller, M.J.,
Kenyon, S.J., and Livio, M. (2010) A Galactic Origin for HE 0437--5439,
The Hypervelocity Star Near the Large Magellanic Cloud. \textit{The
Astrophysical Journal Letters}, 719:1, L23-L27.\vspace{0.5ex}

Bussard, R.W. (1960) Galactic Matter and Interstellar Flight,
\textit{Astronautica Acta}, 6, 179-194.\vspace{0.5ex}

Cox, A.N (2000) \textit{Allen's Astrophysical
Quantities}, Fourth Edition. Springer-Verlag, New York, USA. \vspace{0.5ex}

Dyson, F.J. (1968) Interstellar transport. \textit{Physics Today},
Oct., 41--45.\vspace{0.5ex}

Edmondson, W.H., and Stevens, I.R (2003) The utilization of pulsars
as SETI beacons. \textit{International Journal of Astrobiology}, 2:4,
231--271.\vspace{0.5ex}

Einstein, A. (1905) Zur Elektrodynamik bewegter K\"orper,
\textit{Annalen der Physik}, 322:10, 891--921.\vspace{0.5ex}

Freeman, J.W. (1998), Interplanetary Heliocentric Coordinates for
the Helios 2 Spacecraft, \textit{NSSDC (National Space Science Data
Center)},  available from 
\url{http://nssdcftp.gsfc.nasa.gov/spacecraft_data/helios/helios2/traj/}.
Data-set data set HELIOS\_2/MGD\_FREEMAN. \vspace{0.5ex}

Freitas Jr., R.A. (1983) The Search for Extraterrestrial Artifacts
(SETA). \textit{Journal of the British Interplanetary Society}, 36,
501-506.\vspace{0.5ex}

Forward, R.L. (1985) Antiproton Annihilation Propulsion,
\textit{USAF Rocket Propulsion Laboratory Report AFRPL TR-85-034}. \vspace{0.5ex}

Forward, R.L. (1985) Starwisp: An Ultra-Light Interstellar Probe.
\textit{Journal of Spacecraft}, 22:3, 345-350.\vspace{0.5ex}

Garcia-Escartin, J.C., and Chamorro-Posada, P. (2011) Repeaters in
relativistic communications. \textit{arXiv:1106.4131v1}.\vspace{0.5ex}

Ginsburg, I., Loeb, A., Wegner, G.A. (2012) Hypervelocity Planets
and Transits Around Hypervelocity Stars. \textit{arXiv:1201.1446v1}.\vspace{0.5ex}

Hinshaw, G., Weiland, J. L., Hill, R. S., Odegard, N., Larson, D.,
Bennett, C. L., Dunkley, J., Gold, B., Greason, M. R., Jarosik, N.,
Komatsu, E., Nolta, M. R., Page, L., Spergel, D. N., Wollack, E.,
Halpern, M., Kogut, A., Limon, M., Meyer, S. S., Tucker, G. S., and 
Wright, E. L. (2009) Five-Year Wilkinson Microwave Anisotropy Probe
Observations: Data Processing, Sky Maps, and Basic Results. \textit{The
Astrophysical Journal Supplement Series}, 180:2, 225-245.\vspace{0.5ex}

Hui, C. Y., Becker, W. (2006) Probing the proper motion of the
central compact object in Puppis-A with the Chandra high resolution
camera. \textit{Astronomy \& Astrophysics}, 457:3, L33-L36.\vspace{0.5ex}

Lagoute, C., and Davoust, E. (1995) The interstellar traveler.
\textit{American Journal of Physics}, 63, 221-227.\vspace{0.5ex}

L\'eger, A., Mariotti, J.M., Mennesson, B., Ollivier, M., Puget, J.L.,
Rouan, D., and Schneider, J. (1996) Could we search for primitive life
on extrasolar planets in the near future? The DARWIN project.
\textit{Icarus} 123, 249-255.\vspace{0.5ex}

Loeb, A., and Turner, E.L. (2011) Detection Technique for
Artificially-Illuminated Objects in the Outer Solar System and Beyond.
\textit{arXiv:1110.6181v2}.\vspace{0.5ex}

Long, K.F., Obousy, R.K., and Hein, A. (2011) Project Icarus:
Optimisation of nuclear fusion propulsion for interstellar missions.
\textit{Acta Astronautica}, 68, 1820--1829.\vspace{0.5ex}

Matloff, G.L., and Pazmino, J. (1997) Detecting interstellar
migrations. In \textit{Proceedings of the 5th International Conference on Bioastronomy.}
\textit{IAU Colloquium 161, Astronomical and Biochemical Origins and
the Search for Life in the Universe}. 757-759.\vspace{0.5ex}

Matloff, G.L. (2000) \textit{Deep-Space Probes}. Springer-Praxis,
Chichester, UK.\vspace{0.5ex}

McKinley, J.M. (1979) Relativistic transformations of light power.
\textit{American Journal of Physics}, 47, 602-605.\vspace{0.5ex}

McKinley, J.M. (1980) Relativistic transformation of solid angle.
\textit{American Journal of Physics}, 48, 612-614.\vspace{0.5ex}

McKinley, J.M., and Doherty, P. (1979) In search of the
``starbow'': The appearance of the starfield
from a relativistic spaceship. \textit{American Journal of Physics},
47:4, 309-316.\vspace{0.5ex}

NASA (1969) \textit{Apollo 10 mission report}, MSC-00126, 6-13.\vspace{0.5ex}

NASA (2003) \textit{Galileo End of Mission press kit}, p. 12. \vspace{0.5ex}

Sagan, C., Thompson, W.R., Carlson, R., Gurnett, D., and Hord, C.
(1993) A search for life on Earth from the Galileo spacecraft.
\textit{Nature} 365, 715--721.\vspace{0.5ex}

Semyonov, O.G. (2008) Radiation hazard of relativistic interstellar
flight. \textit{Acta Astronautica}, 64:5-6, 644-653.\vspace{0.5ex}

Tarter, J. (2001) The search for extraterrestrial intelligence
(SETI). \textit{Annual Review of Astronomy and Astrophysics}, 39, 511-548.\vspace{0.5ex}

Turnbull, M. C., and Tarter, J. (2003a) Target selection for SETI.
I. A catalog of nearby habitable stellar systems. \textit{The Astrophysical Journal Supplement}, 145, 181-198.\vspace{0.5ex}

Turnbull, M. C., and Tarter, J. (2003b) Target selection for SETI.
II. Tycho-2 dwarfs, old open clusters, and the nearest 100 stars.
\textit{The Astrophysical Journal Supplement}, 149, 423-436.\vspace{0.5ex}

Vulpetti, G., Johnson, L., and Matloff, G.L. (2008) \textit{Solar
Sails: A Novel Approach to Interplanetary Travel}. Springer-Praxis,
USA.\vspace{0.5ex}

Zubrin, R. (1995) Detection of Extraterrestrial Civilizations via
the Spectral Signature of Advanced Interstellar Spacecraft. \textit{ASP
Conference Series, Progress in the Search for Extraterrestrial Life},
74, 487-497.\vspace{0.5ex}

\end{document}